\documentclass[english,aps,prr,onecolumn,notitlepage,showpacs,superscriptaddress,nofootinbibfloatfix,nofootinbib]{revtex4-1}
\usepackage[T1]{fontenc}
\usepackage[latin1]{inputenc}
\usepackage{graphicx}
\usepackage{bm}
\usepackage{amssymb}
\usepackage{color}
\usepackage[usenames,dvipsnames]{xcolor}
\usepackage{amsmath}
\usepackage{amstext}
\usepackage{latexsym}
\usepackage[colorlinks=true,citecolor=Cerulean,linkcolor=RubineRed,urlcolor=Cerulean]{hyperref}
\usepackage{lipsum}
\usepackage{enumitem}

\usepackage{mathptmx} 
\usepackage{verbatim}

\usepackage{amsfonts}
\usepackage{epsfig}
\usepackage{dsfont}
\usepackage{mathrsfs}
\usepackage{arydshln,leftidx,mathtools}

\begin{document}

\title{Nonadiabatic energy fluctuations of  scale-invariant quantum systems in a time-dependent trap}

\author{Mathieu Beau}
\affiliation{Department of Physics, University of Massachusetts, Boston, MA 02125, USA}
\author{Adolfo del Campo}
\affiliation{Donostia International Physics Center,  E-20018 San Sebasti\'an, Spain}
\affiliation{IKERBASQUE, Basque Foundation for Science, E-48013 Bilbao, Spain}
\affiliation{Department of Physics, University of Massachusetts, Boston, MA 02125, USA}
\affiliation{Theory Division, Los Alamos National Laboratory, MS-B213, Los Alamos, NM 87545, USA}

\def\q{{\bf q}}

\def\G{\Gamma}
\def\L{\Lambda}
\def\la{\lambda}
\def\g{\gamma}
\def\al{\alpha}
\def\s{\sigma}
\def\e{\epsilon}
\def\k{\kappa}
\def\ve{\varepsilon}
\def\l{\left}
\def\r{\right}
\def\te{\mbox{e}}
\def\d{{\rm d}}
\def\t{{\rm t}}
\def\K{{\rm K}}
\def\N{{\rm N}}
\def\H{{\rm H}}
\def\la{\langle}
\def\ra{\rangle}
\def\om{\omega}
\def\Om{\Omega}
\def\vep{\varepsilon}
\def\wh{\widehat}
\def\tr{{\rm Tr}}
\def\da{\dagger}
\def\iz{\left}
\def\zi{\right}
\newcommand{\beq}{\begin{equation}}
\newcommand{\eeq}{\end{equation}}
\newcommand{\beqa}{\begin{eqnarray}}
\newcommand{\eeqa}{\end{eqnarray}}
\newcommand{\intf}{\int_{-\infty}^\infty}
\newcommand{\into}{\int_0^\infty}

\begin{abstract}
We consider the nonadiabatic energy fluctuations of a many-body system  in a time-dependent harmonic trap. In the presence of scale-invariance, the dynamics becomes self-similar and the nondiabatic energy fluctuations can be found in terms of  the initial expectation values of the second moments of the Hamiltonian, square position, and squeezing operators. Nonadiabatic features are expressed in terms of the scaling factor governing the size of the atomic cloud, which can be extracted from time-of-flight images. We apply this exact relation to a number of examples: the single-particle harmonic oscillator, the  one-dimensional Calogero-Sutherland model, describing bosons with inverse-square interactions that includes the non-interacting Bose gas and the Tonks-Girdardeau gas as limiting cases, and the unitary Fermi gas. We illustrate these results for various expansion protocols involving sudden quenches of the trap frequency, linear ramps and shortcuts to adiabaticity. Our results pave the way to the experimental study of nonadiabatic energy fluctuations in driven quantum fluids.

\end{abstract}

\maketitle

\section{Introduction}

Nonequilibrium quantum phenomena in many-body systems are often hard to describe by numerical methods due to the growth of entanglement with the time of evolution. 
In some cases, exact solutions can be found by analytical methods. This is often the case when the dynamics is characterized by the presence of dynamical invariants, also known as  invariants of motion.
Such conserved quantities can give rise to symmetries manifested in the dynamics. 
A prominent example is the presence of scale-invariance in the dynamics of a broad family of quantum fluids in time-dependent traps, which is exploited with routine in ultracold atom laboratories \cite{Dalfovo99,Giorgini08}.
In this context, time-of-flight imaging techniques rely on the  scale-invariant dynamics of a suddenly released atomic cloud, that expands freely  \cite{CastinDum96,Kagan96,CastinDum98}.
Scale invariance is then manifested in  local correlation functions, such as the density profile or the density-density correlation function, that become self-similar. One can then relate the correlations measured at a given time after the expansion with those of the initial trapped system. 

Scale invariance is  not restricted to free dynamics and it is also common in the presence of interactions. Indeed, in harmonically trapped systems it is always realized when the  interparticle potential has the same scaling dimension as the kinetic energy. As an example, Bose-Einstein condensates with s-wave contact interactions exhibit it exactly in two-spatial dimensions as a result of the Pitaevskii-Rosch symmetry \cite{PR97}.
Similarly,  it is present in the one-dimensional (rational) Calogero-Sutherland gas describing bosons with inverse-square interactions \cite{Sutherland98}.
Scale invariance is also manifested in systems with divergent interactions.
In one-dimensional harmonically-confined systems it is respected exactly in the presence of hard-core interactions. It thus rules the dynamics of spin-polarized fermions (free fermions, where the Pauli exclusion principle gives rise to an  effective hard-core interaction) as well as bosons in the Tonks-Girardeau regime \cite{RigolMuramatsu05,MinguzziGangardt05,delcampo08}. 
In three spatial dimensions it governs the dynamics of a unitary Fermi gas \cite{Castin04,Deng16}. Finally, scale-invariance can also be invoked effectively in some regimes. Ultracold bosons well-described by mean-field theory exhibit this symmetry in arbitrary spatial dimensions in the Thomas-Fermi regime,  where the contribution of the kinetic energy  can be neglected \cite{CastinDum96,Kagan96,CastinDum98}.

Beyond the use of time-of-flight imaging techniques, scale invariance has applications in a wide variety of contexts. It allows for the engineering of fast control protocols known as shortcuts to adiabaticity \cite{MasudaNakamura10,Chen10,delcampo11,delcampo13,DeffnerJarzynskiAdC14}, as demonstrated in the laboratory \cite{Schaff10,Schaff11,Rohringer2015,An16,Deng18,Diao18}.
It makes possible an exact treatment of the finite-time thermodynamics \cite{Jaramillo16}  and its optimization \cite{Schmiedl07,Choi2011,Choi2011b,Beau16}, as well as the suppression of quantum friction in quantum machines (like engines and refrigerators) that use a scale-invariant quantum fluid as a working substance \cite{Campo2014a,Beau16,delCampo2018}. Exploiting scale invariance, frictionless expansion and compression strokes have been implemented in a unitary Fermi gas \cite{Deng18Sci}. In addition, the study of scale invariance in the presence of perturbations can be used as a probe for quantum anomalies \cite{Olshanii10}, the finite amplitude of the interaction strength \cite{Zhang14}, and the breakdown of effective descriptions based on dimensional reduction \cite{Merloti13}.
In foundations of physics, it  can shed light on regimes that would be challenging to explore experimentally or numerically.
For instance, it can be used to describe exactly the  quantum decay of  quantum states at arbitrarily long times \cite{delcampo16}.

While local correlations functions are simply related at different times via scale invariance, other observables exhibit a more intricate dependence. A prominent one is the mean energy of the system that has been studied both in single-particle \cite{Lohe08} and many-particle \cite{Jaramillo16,Beau16} quantum systems and that is crucial for applications in quantum thermodynamics.

In this work, we focus on the characterization of energy fluctuations in many-particle quantum systems exhibiting scale invariance. Energy fluctuations are known to govern  characteristic time scales in the dynamics in view of the Mandelstam-Tamm time-energy uncertainty relation \cite{MT45}. More generally, they play a crucial role in the formulation of speed limits, that establish the minimum time for a state to evolve into a distinct state, in both classical \cite{Shanahan18,Okuyama18} and quantum \cite{AA90} systems. Energy fluctuations can also be used to quantify the cost of thermodynamic processes \cite{Demirplak2008,Campbell17}, including shortcuts to adiabaticity, and their minimization paves the way to the finding of optimal protocols, e.g., the solution of the quantum brachistochrone problem \cite{Carlini05,Takahashi13} .


\section{Exact many-body dynamics under scale invariance}

The description of the exact many-body dynamics under scale invariance can be simplified using  a sequence of symmetry transformations. 
To this end, we follow closely Lohe \cite{Lohe08} and in particular the formulation for many-body systems in \cite{Jaramillo16,Beau16}.
Consider a $\N$-particle quantum system  in $D$ spatial dimensions and let us denote by $\vec{r}_i$ and $\vec{p}_i$ the position and momentum of the $i$-th particle, respectively.
We consider a confined  many-particle system described by the time-dependent Hamiltonian
\begin{eqnarray}
 H(t)=\sum_{i=1}^{\N}\left[\frac{\vec{p_i}^2}{2m}+\frac{1}{2}m\omega(t)^2 \vec{r_i}^{2}\right]+\sum_{i<j}V(\vec{r}_i-\vec{r}_j),
 \label{hscale}
\end{eqnarray}
where the interparticle interaction potential  exhibits the  scaling property $V(\lambda \vec{r})=\lambda^{-2}V(\vec{r})$. The dynamical group for the Hamiltonian \eqref{hscale} is $SU(1,1)$ \cite{Gambardella75}. Given an  arbitrary energy eigenstate  at $t=0$, the exact dynamics generated by the  Hamiltonian \eqref{hscale}  for any modulation of the trapping frequency $\om(t)$ is described by a scaling law \cite{Gritsev10,delcampo11,delcampo13}, 
\beqa
\label{psit}
\Psi_{n}\!\left(
\vec{r}_1,\dots,\vec{r}_\N,t\right)\!\!\!=\!
b^{-\frac{D\N}{2}}\exp\left[i\frac{m\dot{b}}{2\hbar b}\sum_{i=1}^\N \vec{r_i}^2-i\int_{0}^t\frac{E_n(0)}{\hbar b(t')^2}dt'\right]\Psi_{n}\left(\frac{\vec{r}_1}{b},\dots,\frac{\vec{r}_\N}{b},t=0\!\right),
\eeqa
where  $ n=\{n_i:i=1,\dots,D\N\}$ is a shorthand for a complete set of quantum numbers that labels the quantum state as well as the corresponding energy eigenvalue  $E_n(0)=E_n[\om(0)]$.
The  scaling factor  $ b=b(t)>0$  as a function of time is the solution of the Ermakov  equation
\beqa
\label{EPE}
\ddot{b}+\om(t)^2b=\om_0^2b^{-3},
\eeqa
with the initial  boundary conditions $b(0)=1$ and $\dot{b}(0)=0$, such that the  time-dependent state (\ref{psit}) reduces to the initial eigenstate at the beginning of the process $t=0$.

The time-dependent state (\ref{psit}) can be obtained by acting with a sequence of unitary transformations on the initial state.
We follow Lohe \cite{Lohe08} and  Jaramillo et al. \cite{Jaramillo16} to describe the time-evolution in terms of the action of two  unitary transformations, elements of $SU(1,1)$. To this end, we introduce 
\beqa
T_{\rm dil}&=&\exp\left[-i\frac{\log b}{2\hbar}\sum_{i=1}^\N (\vec{r}_i\cdot\vec{p}_i+\vec{p}_i\cdot\vec{r}_i)\right],\\
T_r&=&\exp\left[i\frac{m\dot{b}}{2\hbar b}\sum_{i=1}^\N \vec{r_i}^2\right].
\eeqa
We note that $T_{\rm dil}$ implements scaling transformations by acting on a function in the coordinate representation  $f(\vec{r}_1,\dots,\vec{r}_\N)$ according to $T_{\rm dil}f(\vec{r}_1,\dots,\vec{r}_\N)T_{\rm dil}^\dag=f(\vec{r}_1/b,\dots,\vec{r}_\N/b)$. Similarly, acting on a function in the momentum representation $g(\vec{p}_1,\dots,\vec{p}_\N)$, one finds 
$T_{\rm dil}g(\vec{p}_1,\dots,\vec{p}_\N)T_{\rm dil}^\dag=g(\vec{p}_1b,\dots,\vec{p}_\N b)$.
By contrast, $T_r$ acts as a multiplicative factor in the coordinate representation while it generates translations in momentum space acting on a function $g(\vec{p}_1,\dots,\vec{p}_\N)$ as $T_rg(\vec{p}_1,\dots,\vec{p}_\N)T_r^\dag=g(\vec{p}_1-m\dot{b}\vec{r}_1/b,\dots,\vec{r}_\N-m\dot{b}\vec{r}_\N/b)$.

Denoting $T=T_rT_{\rm dil}$ and  the Hamiltonian \eqref{hscale} at $t=0$ by $H_0$, we identify the invariant of motion 
\beqa
\label{mbinv}
\mathcal{I}&=&TH_0T^{\dag}\nonumber\\
&=&\sum_{i=1}^{\N}\left[\frac{b^2\vec{p_i}^2}{2m}+\frac{1}{2}m\vec{r_i}^{2}\left(\frac{\omega(0)^2}{b^2}+\dot{b}^2\right)
-\frac{b\dot{b}}{2}\ (\vec{r}_i\cdot\vec{p}_i+\vec{p}_i\cdot\vec{r}_i)\right]
+b^2\sum_{i<j}V(\vec{r}_i-\vec{r}_j),
\eeqa
as an instance of the many-body invariants of motion known under scale-invariant driving \cite{delcampo13,DeffnerJarzynskiAdC14}. Equation (\ref{mbinv}) is indeed a generalization of the Lewis-Riesenfeld invariant of motion known for a single-particle driven harmonic oscillator \cite{LewisRiesenfeld69}, satisfies 
\beqa
\frac{\partial}{\partial t}\mathcal{I}=\frac{1}{i\hbar}[H(t),\mathcal{I}], 
\eeqa
and has time-dependent eigenfunctions $\Phi_n\left(\vec{r}_1,\dots,\vec{r}_\N,t\right)=T\Psi_{n}\left(\vec{r}_1,\dots,\vec{r}_\N,0\right)$ and time-independent eigenvalues $E_n(0)$. In terms of the eigenvectors of  $\mathcal{I}$  the time-evolving state \eqref{psit} can be written as 
\begin{equation}\label{psit2}
\Psi_{n}\left(\vec{r}_1,\dots,\vec{r}_\N,t\right)=e^{i\alpha_t}T\Psi_{n}\left(\vec{r}_1,\dots,\vec{r}_\N,0\right),
\end{equation}
where $\alpha_t=-\int_{0}^t\frac{E_n(0)}{\hbar b(t')^2}dt'$ is a time-dependent dynamical phase.

\section{Exact nonadiabatic mean energy}

Knowledge of the exact time-dependent dynamics under scale invariance makes possible to find the exact  nonadiabatic mean energy of the system, as shown in \cite{Jaramillo16}. 
To simplify the notation, we introduce the square position and squeezing operators
\begin{equation}
Q =\sum_{i=1}^{\N}\vec{r_i}^2,\quad 
C =\frac{1}{2}\sum_{i=1}^{\N}(\vec{r}_i\cdot\vec{p}_i+\vec{p}_i\cdot\vec{r}_i)=\frac{1}{2}\sum_{i=1}^{\N}\left\{\vec{r}_i,\vec{p}_i\right\},
\end{equation}
 and note that the time-dependent Hamiltonian (\ref{hscale}) can be rewritten as
\beqa
H(t)=\frac{1}{b^2}\mathcal{I}+i\hbar\frac{\partial T}{\partial t}T^{\dag},
\eeqa 
where
\beqa
i\hbar\frac{\partial T}{\partial t}T^{\dag}=\left[-\frac{1}{2}m\left(\frac{\ddot{b}}{b}+\frac{\dot{b}^2}{b^2}\right)Q+\frac{\dot{b}}{b}C\right].
\eeqa
Using Eq.  \eqref{psit2}, the expectation value of the time-dependent Hamiltonian in the time-evolving state reads 
\beqa
\la\Psi_{n}(t)|H(t)|\Psi_{n}(t)\ra=
\la\Psi_{n}(0)|T^\dagger H(t) T|\Psi_{n}(0)\ra
=\la\Psi_{n}(0)|\left(\frac{H_0}{b^2}+i\hbar T^\dagger \frac{\partial T}{\partial t}\right)|\Psi_{n}(0)\ra,
\eeqa
with 
\beqa
i\hbar\ T^\dagger \frac{\partial T}{\partial t}=\left[\frac{m}{2}\left(\dot{b}^2-\ddot{b}b\right)Q+\frac{\dot{b}}{b}C\right].
\eeqa
As a result, the mean nonadiabatic energy is given by
\beqa
\la\Psi_{n}(t)|H(t)|\Psi_{n}(t)\ra=
\frac{E_n(0)}{b^2}-\frac{m}{2}\left(b\ddot{b}
-\dot{b}^2\right)\la\Psi_{n}(0)|Q|\Psi_{n}(0)\ra+\frac{\dot{b}}{b}\la\Psi_{n}(0)| C|\Psi_{n}(0)\ra.
\eeqa
As this expression holds for an arbitrary eigenstate, the expectation value can be also carried out with respect to an arbitrary state diagonal in the energy eigenbasis, such as an equilibrium thermal state. Specifically, the nonadiabatic mean-energy following a change of $\om(t)$ is found to be 
\beqa
\label{avenscaling}
 \la H(t)\ra_t =\frac{1}{b^2} \la H(0)\ra+\frac{m}{2}(\dot{b}^2-b\ddot{b})\la Q(0)\ra\ +\frac{\dot{b}}{b}\la C(0)\ra,
\eeqa
 where $\la \bullet \ra \equiv \text{tr}\left(\bullet \rho_0\right) $ denotes the mean value at the time $t=0$ with respect to a state $\rho_0$ diagonal in the eigenbasis of $H(0)$. Similarly,  $\la \bullet \ra_t \equiv \text{tr}\left(\bullet \rho_t\right) $ denotes the mean value at the time $t$ with respect to the state $\rho_t=U(t)\rho_0U(t)^\dag$, where $U(t)$ is the time evolution operator. Under scale invariance it is thus possible to relate the evolution of the mean nonadiabatic energy to the initial expectation values of the Hamiltonian,  square position and squeezing operators. \\

For an initial state at equilibrium, diagonal in the eigenbasis of $H(0)$,  the mean squeezing vanishes $\la C(0)\ra=0$. 
In addition, the initial expectation value of $Q$ is proportional to the mean energy of the system \cite{Jaramillo16},
\begin{equation}\label{meansquareformula}
\la Q(0)\ra = \frac{\la H(0)\ra}{m\omega_0^2}.
\end{equation}
Using this property and the Ermakov equation (\ref{avenscaling}), the nonadiabatic  mean energy reads
\beqa
\la H(t)\ra_t=Q^{\ast}(t)\frac{\om(t)}{\om_0} \la H(0)\ra=Q^{\ast}(t)\la H(t)\ra_{\rm ad},\label{aven}
\eeqa
where we have introduced the adiabatic mean energy $\la H(t)\ra_{\rm ad}= \la H(0)\ra\om(t)/\om(0)$ as well as the  nonadiabatic factor $Q^{\ast}(t)$ that accounts for the amount of energy excitations over the adiabatic dynamics
\beqa\label{Qstar}
Q^{\ast}(t)&=&\frac{\om_0}{\om(t)} \left[\frac{1}{b(t)^2}+\frac{\dot{b}(t)^2-b(t)\ddot{b}(t)}{2\om_0^2}\right]
=\frac{\om_0}{\om(t)} \left[\frac{1}{2b(t)^{2}}+\frac{\om(t)^2 b(t)^2}{2\om_0^2}+\frac{\dot{b}(t)^2}{2\om_0^2}\right].
\eeqa
Equations \eqref{aven} and \eqref{Qstar} provide the exact nonadiabatic mean energy of the family of systems governed by Hamiltonian \eqref{hscale}.  We emphasize that the nonadiabatic factor $Q^{\ast}(t)$ is simply the ratio between the nonadiabatic and adiabatic mean energies. As it can be expressed in terms of the scaling factor, that governs the size of the atomic cloud, it is an experimentally accessible quantity \cite{Deng18Sci,Diao18}.

Summarizing these background results, reported in \cite{Jaramillo16},  it suffices to know the evolution of the scaling factor and the initial expectation values at $t=0$ of the mean energy,   square position and squeezing operators to determine the exact nonadiabatic dynamics of the energy fluctuations. This relation is particularly simple when the initial state is at equilibrium and the mean squeezing vanishes. As we shall see in the next section, a similar exact relation can be derived for the  second order of the Hamiltonian required to describe nonadiabatic energy fluctuations, but requires the computation of expectation values at zero time which are model dependent (we have not been able to relate them to properties of the Hamiltonian via symmetries).

\section{Nonadiabatic  energy fluctuations}
We next turn to the focus of our work, the characterization of the energy fluctuations.
To this end, we consider the expectation value of the second momentum of the Hamiltonian at time $t>0$ in the corresponding time-evolving state 
\begin{equation}\label{MeanH(T)^2}
\la \Psi_{n}(t) |H(t)^2| \Psi_{n}(t) \ra = \la \Psi_{n}(0) | T^\dagger H(t)^2 T| \Psi_{n}(0) \ra
= \la \Psi_{n}(0) |( T^\dagger H(t)T )^2| \Psi_{n}(0) \ra.
\end{equation}
We recall that 
\beqa
T^\dagger H(t)T=\frac{H_0}{b^2}+i\hbar T^\dagger \frac{\partial T}{\partial t}=\frac{H_0}{b^2}+\alpha Q+\beta C,
\eeqa
where for compactness we define the time-dependent coefficients
\begin{equation}
\alpha =\frac{m}{2}\left(\dot{b}^2-\ddot{b}b\right), \quad 
 \beta =\frac{\dot{b}}{b}.
\end{equation}
We thus find 
\begin{eqnarray}\label{THTsquare}
( T^\dagger H(t)T )^2 =\frac{H_0^2}{b^4}+\alpha^2 Q^2+\beta^2 C^2 + \frac{\alpha}{b^2}\left\{H_0,Q\right\} + \frac{\beta}{b^2}\left\{H_0,C\right\} + \alpha\beta\left\{Q,C\right\}.
\end{eqnarray}

Evaluation of the expectation value of the fourth, fifth and the sixth terms in \eqref{THTsquare} over an energy eigenstate, or any initial state $\rho_0$ commuting with $H_0$, is straightforward and yields
\begin{subequations}\label{456MeanH(T)^2}
\begin{equation}\label{4MeanH(T)^2}
\la\left\{H_0,Q\right\}\ra=\frac{2\la H_0^2\ra }{m\om_0^2},
\end{equation}
\begin{equation}\label{5MeanH(T)^2}
\la\left\{H_0,C\right\}\ra=2\la H_0\ra \la C\ra = 0,\
\end{equation}
\begin{equation}\label{6MeanH(T)^2}
\la\left\{Q,C\right\}\ra=\frac{im}{2\hbar}\la\left\{Q,[H_0,Q]\right\}\ra
=\frac{im}{2\hbar}\la[H_0,Q^2]\ra=0,
\end{equation}
\end{subequations}
as $\la Q \ra = \la H_0 \ra/m\om_0^2$ from equipartition theorem and $\la C \ra = 0$ in an equilibrium eigenstate. 
Although the mean value of the squeezing operator vanishes, the analogue quantity for its square is generally non-zero. In this case, one needs to specify the model to compute explicitly these quantities. 
Nonadiabatic energy fluctuations are set by the second moment of the Hamiltonian that equals
\begin{equation}\label{MeanHsquare}
\fbox{$\la H(t)^2 \ra_t \equiv \text{tr}\left(\rho(t) H(t)^2\right) =  \la H_0^2\ra\left(\frac{1}{b^4}+\frac{2\alpha}{m\om_0^2b^2}\right) + \alpha^2 \la Q^2\ra +\beta^2 \la C^2\ra$}
\end{equation}
where we remind that  $\la \bullet \ra \equiv \text{tr}\left(\bullet \rho_0\right)$  denotes the mean value at  time $t=0$. This is the main result of this work. As for the mean energy, the nonadiabatic energy fluctuations can be related to the initial expectation value of the second moment of the Hamiltonian, square position and squeezing operators. 
Once these quantities are known, nonadiabatic effects are simply encoded in the time-dependent scaling factor $b(t)$, as the coefficients $\alpha(t)$and $\beta(t)$ are directly determined by it.

At this stage we are ready to consider the adiabatic evolution, corresponding to the limit $\dot{\om
}/\om^2\ll 1$ \cite{LewisRiesenfeld69}. At the level of the Ermakov equation, one can set $\dot{b}\approx0$ and $\ddot{b}\approx 0$ and find that the scaling factor in the adiabatic limit is set by $b_{\rm ad}(t)=[\om(0)/\om(t)]^{1/2}$. As a result, 
$\alpha\approx\beta\approx0$ and the adiabatic  evolution yields

\beqa
\la H(t)^2 \ra_t=\frac{\la H(0)^2 \ra}{b_{\rm ad}^4}=\frac{\om(t)^2}{\om(0)^2}\la H(0)^2 \ra.
\eeqa

\section{Nonadiabatic energy fluctuations: Explicit Examples}

\subsection{Single-particle time-dependent quantum harmonic oscillator} 

Equation \eqref{MeanHsquare} determines exactly  non-equilibrium energy fluctuation as a function of the initial expectation values at time $t=0$. However, it is not possible in general to obtain a simple expression similar to \eqref{aven} for the mean value of the non-equilibrium energy. In \eqref{5MeanH(T)^2} we use that the mean value of the squeezing operator vanishes, but this is not true for its second moment. This makes the explicit computation model dependent and challenging, in general.     
To illustrate this observation, let us evaluate $\la n| C^2 |n \ra\equiv \la \Psi_n(0)| C^2 |\Psi_n(0) \ra$ for a single particle in a harmonic oscillator that corresponds to $H_0$ for $\N=1$, with Hamiltonian $H_0=p^2/2m+m\om_0^2x^2/2$. 
Using the creation and annihilation operators, one finds
\begin{equation}\label{C^2}
\la n| C^2 |n \ra =\frac{\hbar^2}{2}(n^2+n+1), \quad 
\la n| Q^2 |n \ra = \left(\frac{\hbar}{2m\om_0}\right)^2(6n^2+6n+3).
\end{equation}

The nonadiabatic expectation value of the second moment of the Hamiltonian then reads  
\begin{eqnarray}\label{H_n}
\la \Psi_n(t) | H(t)^2 | \Psi_n(t) \ra 
&=&E_n(t)^2 +\frac{\hbar^2}{16\om_0^2}(6n^2+6n+3)\left[\left(\dot{b}^2+\om(t)^2b^2-\frac{\om_0^2}{b^2}\right)^2+\frac{4\om_0^2\dot{b}^2}{b^2}\right]\nonumber\\
&=& E_n(t)^2 +\frac{\hbar^2}{4}(6n^2+6n+3)\left[\left(\om(t)Q^*-\frac{\om_0}{b^2}\right)^2+\frac{\dot{b}^2}{b^2}\right]\nonumber\\
&=& E_n(t)^2 +\frac{\hbar^2\om(t)^2}{4}(6n^2+6n+3)\left[\left(Q^*\right)^2-1\right]\nonumber\\
&=&\la H_t\ra_t^2+\frac{\hbar^2\om(t)^2}{2}(n^2+n+1)\left[\left(Q^*\right)^2-1\right],
\end{eqnarray}
where $Q^*$ is given by Eq. (\ref{Qstar}),  $E_n(t)=\hbar\omega(t)(n+\frac{1}{2})$ is the  instantaneous energy eigenvalue at time $t$ and $\la H_t\ra_t=\la \Psi_n(t) | H(t)| \Psi_n(t) \ra =Q^*E_n(t)$ the corresponding  nonadiabatic mean energy.
The noandiabatic energy variance $\Delta_n H(t)^2 \equiv \la \Psi_n(t) | H(t)^2 | \Psi_n(t) \ra - \la \Psi_n(t) |H(t) |\Psi_n(t)\ra^2$ in a given eigenmode is therefore
\begin{eqnarray}\label{DeltaH_n}
\Delta_n H(t)^2 =\frac{\hbar^2\om(t)^2}{2}(n^2+n+1)\left[\left(Q^*\right)^2-1\right].
\end{eqnarray}
Using (\ref{MeanHsquare}), our result reproduces the prediction by Husimi \cite{Husimi53} and differs from that by Chen and Muga \cite{ChenMuga10} (as the latter result does not vanish in the adiabatic limit, it is incorrect).
We verify that in the adiabatic limit when $\dot{b}=0$, $b=b_{\rm ad}$ and  $Q^*=1$, energy fluctuations in a single driven energy mode vanish identically, as expected.

For a canonical thermal state as initial state, with density matrix $\rho_0=e^{-\beta H(0)}/Z_0$ normalized in terms of the partition function $Z_0={\rm tr}[e^{-\beta H(0)}]=(2\sinh[\beta\hbar\om_0/2])^{-1}$, one finds
\beqa
\Delta H(t)^2 =\frac{\hbar^2\om(t)^2}{4}\frac{1}{\sinh[\beta\hbar\om_0/2]^2}\left[(Q^*)^2+\left(\left(Q^*\right)^2-1\right)\cosh[\beta\hbar\om_0]\right].
\eeqa

\subsection{Calogero-Sutherland gas in a time-dependent harmonic trap} 

We next consider the nonadiabatic energy fluctuations in a driven many-body quantum system.
We focus on  the (rational) Calogero-Sutherland model describing one-dimensional bosons subject to inverse-square pairwise interactions \cite{Calogero71,Sutherland71}. The system Hamiltonian is
\beqa
 H(t)=\sum_{i=1}^{\N}\left[-\frac{\hbar^2}{2m}\frac{\partial^2}{\partial z_i^2}+\frac{1}{2}m\omega(t)^2 z_{i}^{2}\right]+\frac{\hbar^2}{m}\sum_{i<j}\frac{g(g-1)}{(z_i-z_j)^2}.
\eeqa
In one dimension, interactions are inextricably linked to particle statistics. The notion of generalized exclusion statistics  \cite{Haldane91,Wu94}, complementary to the familiar exchange statistics, classifies particles according to their exclusion in Hilbert space.  In this context, the Calogero-Sutherland model can be described as a non-interacting gas of particles obeying generalized exclusion statistics \cite{MurthyShankar94}. The constant $g\geq 0$ thus plays a dual role. While in the standard description it determines  the strength of the inverse-square interactions,  it also serves to parameterize generalized exclusion statistics.  The value $g=0$ corresponds to ideal bosons. 
The case $g=1$ corresponds to hard-core bosons in the continuum, i.e., the so-called Tonks-Girardeau gas \cite{Girardeau60}.
For $g=1/2$ the system is sometimes referred to as composed of Haldane semions, and more generally particles are called geons for an arbitrary value $g$ of the exclusion parameter.
The many-body wavefunction  of the Calogero-Sutherland ground-state at $t=0$ reads
\beqa
\label{gsCSM}
\Psi_0\left(z_1,\dots,z_\N\right)=C_0\prod_{i<k}|z_j-z_i|^{g} \prod_{i=1}^\N\exp\left(-\frac{m\om_0}{2\hbar}z_i^2\right),
\eeqa
where $C_0$ is a known normalization constant, see e.g. \cite{delcampo16}, and the corresponding energy eigenvalue reads $E_0=\hbar\om_0\N[1+g(\N-1)]/2$.
Obviously, for $g=0$ on recovers the wavefunction  an ideal Bose gas, while for $g=1$ one recognizes the wavefunction of bosons in the Tonks-Girardeau regime \cite{Girardeau01}.
The time evolution of  (\ref{gsCSM})  exhibits scale invariance  \cite{Sutherland98,delcampo16} and takes the form of  Eq. (\ref{psit}) in one spatial dimension with $D=1$.
As the ground state is an energy eigenstate,  the expectation value of the second moment of the Hamiltonian simply reads $\la H_0^2\ra=E_0^2$.
For the computation of $\la Q^2\ra$ and  $\la C^2\ra$ we note that these moments can be derived from the corresponding characteristic function, 
which can be written down as an overlap between ground-state wavefunctions. In particular, as the  ground-state is a homogeneous function times a Gaussian term, the
explicit computation of the overlap is possible \cite{delcampo16,Ares18}.

We first  proceed to the computation of $\la \hat{Q}^2\ra$ by introducing the overlap
\beqa
A_Q(\lambda)=\la \Psi_0|e^{-\lambda \hat{Q}}|\Psi_0\ra=
\left[1+\lambda\frac{\hbar}{m\om_0}\right]^{-\frac{\N}{2}[1+g(\N-1)]},
\eeqa
from which moments of $\hat{Q}$ are found as
\beqa
\la \Psi_0|\hat{Q}^k|\Psi_0\ra=(-1)^k\frac{d^k}{d\lambda^k}A_Q(\lambda)\big|_{\lambda=0}.
\eeqa
Their explicit expressions are given by
\beqa
\label{QnCSM}
\la \Psi_0|\hat{Q}|\Psi_0\ra&=&\left(\frac{\hbar}{m\om_0}\right)c,\\
\la \Psi_0|\hat{Q}^2|\Psi_0\ra&=&\left(\frac{\hbar}{m\om_0}\right)^2c(1+c),\dots\\
\la \Psi_0|\hat{Q}^k|\Psi_0\ra&=&\left(\frac{\hbar}{m\om_0}\right)^k\prod_{j=1}^k(c+j-1),
\eeqa
where for compactness we have defined $c=\frac{\N}{2}[1+g(\N-1)]$. 

Similarly, we can evaluate the term $\la C^2\ra$.
To this end, we first notice that the dilation operator acts on a many-body wavefunction as
\beqa
T_{\rm dil}\Psi\left(z_1,\dots,z_\N\right)=b^{-\frac{\N}{2}}\Psi\left(\frac{z_1}{b},\dots,\frac{z_\N}{b}\right).
\eeqa
We next evaluate the overlap
\beqa
A_C(b)=\la \Psi_0|T_{\rm dil}\Psi_0\ra=
\left[\frac{b}{2}\left(1+\frac{1}{b^2}\right)\right]^{-c},
\eeqa
which can be understood as the characteristic function of the eigenvalue distribution of the squeezing operator.
Thus, moments of the generator can  be found using the identity
\beqa
\la \Psi_0|\left[\sum_{i=1}^\N \frac{1}{2}(z_ip_i+p_iz_i)\right]^k|\Psi_0\ra
&=&\left(-i\hbar \right)^k\left(b\frac{d}{db}\right)^kA_0(b)\big|_{b=1}. 
\eeqa
As $A_C(b)$ is a real function, this implies that odd moments $\la \Psi_0|\hat{C}^k|\Psi_0\ra$ identically vanish in the ground state $|\Psi_0\ra$.
By explicit evaluation, we find
\beqa
\la \Psi_0|\hat{C}^2|\Psi_0\ra&=&\hbar^2c,\\
\la \Psi_0|\hat{C}^4|\Psi_0\ra&=&\hbar^4 c(2+3c),\\
\la \Psi_0|\hat{C}^6|\Psi_0\ra&=& \hbar^6c(16+30c+15c^2),\dots
\eeqa

Thus, the second moment of the Hamiltonian is set by
\beqa
 \la H_t^2\ra_t
 &=& E_0^2\left(\frac{1}{b^4}+\frac{2\alpha}{m\om_0^2b^2}\right)+ \alpha^2 \left(\frac{\hbar }{m\om_0}\right)^2c(1+c) +\beta^2 \hbar^2c \nonumber\\
 &=&\hbar^2\om(t)^2c^2\left(Q^*\right)^2+\hbar^2\om(t)^2c\left[\left(Q^*\right)^2-1\right],
 \eeqa
where  $Q^*$ is given by Eq. (\ref{Qstar}) and $E_0=\hbar\om_0c$. As a result, the variance simply reads
\begin{eqnarray}\label{DeltaH_n}
\Delta H(t)^2 =\hbar^2\om(t)^2c\left[\left(Q^*\right)^2-1\right],
\end{eqnarray}
and scales at most quadratically with the number of particles \cite{delcampo16}.

 The exact many-body wavefunctions of the Calogero-Sutherland model are known for excited states and can be written as the product of the ground-state wavefunction and a combination of Hermite polynomials, see e.g. \cite{Vacek94}. Due to the later, the wavefunction of an arbitrary excited state is no longer a homogeneous function times a Gaussian term, precluding the computation of the energy fluctuations  via the method put forward.

 \subsection{Unitary Fermi gas in a time-dependent harmonic trap}
 
We next consider a strongly interacting degenerate Fermi gas with two spin components \cite{Castin12}. The unitary limit is reached when the interaction strength between atoms becomes divergent, e.g., at a Feshbach resonance. To characterize the spectrum and eigenstates of the Hamiltonian, the interactions can be replaced by a contact condition in terms of a function $A$ such that
\beqa
\psi(\vec{r}_1,\dots,\vec{r}_\N)=\frac{A(\vec{R}_{ij},\{\vec{r}_k:k\neq i,j\})}{r_{ij}}+\mathcal{O}(r_{ij}),
\eeqa
where $\vec{R}_{ij}=(\vec{r}_i+\vec{r}_j)/2$ is the center of mass position for particles $i$ and $j$, and the limit of vanishing relative distance between two particles $r_{ij}=|\vec{r}_i-\vec{r}_j|\rightarrow 0$ is taken.
Tan, Werner and Castin have shown that the low-energy wavefunctions of a unitary Fermi gas in an isotropic trap are the product of a Gaussian factor times a homogeneous function \cite{tan2004short,WernerCastin06}. This result together with the scale invariance exhibited by this system \cite{Castin04} will make possible the exact computation of the nonadiabatic energy fluctuations.
 
In three spatial dimensions, we introduce the $3\N$ vector $\vec{X}=(\vec{r}_1,\dots,\vec{r}_\N)$. Similarly, we denote by $\vec{P}=(\vec{p}_1,\dots,\vec{p}_\N)$ the $3\N$ vector of the conjugated momenta.
Denoting the norm of $\vec{X}$ by $X=\sqrt{\sum_j r_j^2}$ and the unit vector $\hat{n}=\vec{X}/X$, we make use of the hyperspherical coordinates $(X,\hat{n})$.
Eigenstates in a trap can be related to zero-energy eigenstates in free space. We follow Werner and Castin \cite{WernerCastin06} and denote by $\psi_\nu^0(\vec{X})$ a zero-energy eigenstate in free space. The latter exhibits  scale invariance
\beqa
\psi_\nu^0(\vec{X}/b)=b^{-\nu}\psi_\nu^0(\vec{X}), 
\eeqa
where $\nu$ is the degree of homogeneity, i.e., $\vec{X}\cdot\vec{P}\psi_\nu=-i\hbar\nu\psi_\nu$ \cite{tan2004short}.
The corresponding trapped eigenstate  reads
\beqa
\psi_\nu(\vec{X})=e^{-\frac{m\om_0X^2}{2\hbar}}\psi_\nu^0(\vec{X})
\eeqa
and its energy eigenvalue is given by 
\beqa
E(0)=\hbar\om_0(\nu+2q-3\N/2),
\eeqa
in terms of a non-negative integer $q$.
Energy eigenstates are connected by raising and lowering operators associated with the $SO(2,1)$ Pitaevskii-Rosch symmetry.
The lowest energy eigenstate of a given tower of states, that with $q=0$ that is annihilated by the lowering  operator, can be shown to take the form
 \beqa
\label{psilowE}
\psi_\nu(\vec{X})=Ce^{-\frac{m\om_0X^2}{2\hbar}}X^{\frac{E}{\hbar\om}-\frac{3\N}{2}}f(\hat{n}),
\eeqa
where $C$ is a normalization constant. This completes our review of \cite{tan2004short,WernerCastin06}.

By scale invariance, the time-dependent state following a modulation of the trapping frequency is of the form in Eq. (\ref{psit}).
As it turns out, this suffices to compute the nonadiabatic expectation value of the first and second energy moments of the Hamiltonian of a unitary Fermi gas  initially prepared in the low-energy state (\ref{psilowE}). To this end, we note that the characteristic functions of $Q=X^2$ and $C=(\vec{X}\cdot\vec{P}+\vec{P}\cdot\vec{X})/2$ are
\beqa
A_Q(\lambda)&=&\la \psi_\nu|e^{-\lambda Q}|\psi_\nu\ra =\left[1+\lambda\frac{\hbar}{m\om_0}\right]^{-\frac{E(0)}{\hbar\om_0}},\\
A_C(b)&=&\la  \psi_\nu|T_{\rm dil} \psi_\nu\ra=\left[\frac{b}{2}\left(1+\frac{1}{b^2}\right)\right]^{-\frac{E(0)}{\hbar\om_0}},
\eeqa
 and thus
 \beqa
 \la  \psi_\nu|\hat{Q}^2| \psi_\nu\ra&=&\left(\frac{\hbar}{m\om_0}\right)^2\frac{E(0)}{\hbar\om_0}\left(1+\frac{E(0)}{\hbar\om_0}\right)\, ,\\
 \la  \psi_\nu|\hat{C}^2| \psi_\nu\ra&=&\hbar^2\frac{E(0)}{\hbar\om_0},
 \eeqa
whence it follows
 \beqa
 \la H_t^2\ra_t=E(0)^2\left[\frac{1}{b^4}+\frac{(\dot{b}^2-\ddot{b}b)}{\om_0^2b^2}\right]+\frac{\hbar^2(\dot{b}^2-\ddot{b}b)^2}{4\om_0^2} \frac{E(0)}{\hbar\om_0}\left(1+\frac{E(0)}{\hbar\om_0}\right) 
 +\hbar^2 \left(\frac{\dot{b}}{b}\right)^2\frac{E(0)}{\hbar\om_0}.
 \eeqa
 
It is convenient to rewrite this expression in the form
\beqa
 \la H_t^2\ra_t=\hbar^2\om(t)^2\left(\frac{E(0)}{\hbar\om_0}\right)^2+\hbar^2\frac{E(0)}{\hbar\om_0}\left(1+\frac{E(0)}{\hbar\om_0}\right)\left[\left(\frac{\dot{b}}{b}\right)^2+\frac{1}{4\om_0^2}\left(\dot{b}^2+\om(t)^2b^2-\frac{\om_0^2}{b^2}\right)^2\right],
\eeqa
in parallel with the results of the single-particle harmonic oscillator as well as that of the Calogero-Sutherland model with the corresponding ground states as initial state.
This last expression can be rewritten as follows
\beqa
 \la H_t^2\ra_t&=&\hbar^2\om(t)^2\left(\frac{E(0)}{\hbar\om_0}\right)^2+\hbar^2\om(t)^2\frac{E(0)}{\hbar\om_0}\left(1+\frac{E(0)}{\hbar\om_0}\right)
 \left[\left(Q^*\right)^2-1\right]\nonumber\\
 &=&\hbar^2\om(t)^2\left(\frac{E(0)}{\hbar\om_0}\right)^2\left(Q^*\right)^2+\hbar^2\om(t)^2\frac{E(0)}{\hbar\om_0}
 \left[\left(Q^*\right)^2-1\right],
\eeqa
with  $Q^*$  given by Eq. (\ref{Qstar}),  leading to the expression of the energy variance  in the form
\beqa
\la H_t^2\ra_t- \la H_t\ra_t^2=\hbar^2\om(t)^2\frac{E(0)}{\hbar\om_0}
 \left[\left(Q^*\right)^2-1\right].
\eeqa
The expression of the results in terms of the ratio $\frac{E(0)}{\hbar\om_0}$ is particularly convenient and it also holds for the results derived in the previous subsections for the ground states of the harmonic oscillator and the  Calogero-Sutherland model. As a result, we shall use it hereafter without specifying the system under consideration. Beyond these models, our results hold whenever the initial state is the product of a Gaussian term times a homogeneous function. 
We further note that
\beqa
\frac{E(0)}{\hbar\om_0}=\la  \psi_\nu|\hat{Q}| \psi_\nu\ra\left(\frac{m\om_0}{\hbar}\right)=\frac{1}{x_0^2}\la  \psi_\nu|X^2| \psi_\nu\ra=\sigma^2,
\eeqa
 which is but the initial size of the cloud in units of $x_0=\sqrt{\hbar/(m\om_0)}$, and is thus an experimentally measurable quantity.

\section{Nonadiabatic moments of the square position operator} 
Due to the $SU(1,1)$ dynamical symmetry group, we have related moments of the Hamiltonian $H(t)$ to those of the operators $H(0)$, $Q$ and $C$ at $t=0$.
For completeness, we note as well the nonadiabatic evolution of the moments of the square position operator $Q$. 
To this end, we introduce
\beqa
A_Q(\lambda,t)=\la \Psi_0(t)|e^{-\lambda \hat{Q}}|\Psi_0(t)\ra=\la \Psi_0(0)|e^{-\lambda T^\dag\hat{Q}T}|\Psi_0(0)\ra.
\eeqa
Noticing that $ T^\dag\hat{Q}T= T_{\rm dil}^\dag\hat{Q}T_{\rm dil}=Qb^2$, one finds
\beqa
A_Q(\lambda,t)=\left[1+\lambda\frac{\hbar b^2}{m\om_0 }\right]^{-\frac{E(0)}{\hbar\om_0}}.
\eeqa
As expected, the $k$-th momentum  of the position operator scales in a trivial form,
\beqa
\la \Psi_0(t)|\hat{Q}^k|\Psi_0(t)\ra=b^{2k}\la \Psi_0(0)|\hat{Q}^k|\Psi_0(0)\ra,
\eeqa
and is proportional to the corresponding moment at $t=0$, in agreement with the definition of the scaling factor $b(t)$.

\section{Nonadiabatic moments  of the squeezing operator}
The evolution of the squeezing operator $C$ is more involved than that of $Q$, and as with the non-adiabatic energy, we shall only compute the first two moments explicitly.
We first note that
\beqa
 T^\dag CT=T_{\rm dil}^\dag \frac{1}{2}\sum_{i=1}^{\N}\left\{\vec{r}_i,\vec{p}_i+\frac{m\dot{b}}{b}\vec{r}_i\right\}T_{\rm dil}=C+m\dot{b}bQ.
\eeqa
The first  moment reads
\beqa
\la \Psi_0(t)|C|\Psi_0(t)\ra=\la\Psi_0(0)|\left(C+m\dot{b}bQ\right)|\Psi_0(0)\ra=m\dot{b}b\la\Psi_0(0)|Q|\Psi_0(0)\ra=\frac{\hbar}{\om_0}\dot{b}b\frac{E(0)}{\hbar\om_0}.
\eeqa
while the second one is given by

\beqa
 \la \Psi_0(t)|C^2|\Psi_0(t)\ra&=&\la\Psi_0(0)|\left(C+m\dot{b}bQ\right)^2|\Psi_0(0)\ra\nonumber\\
 &=&\la\Psi_0(0)|C^2|\Psi_0(0)\ra+m^2\dot{b}^2b^2\la\Psi_0(0)|Q^2|\Psi_0(0)\ra+m\dot{b}b\la\Psi_0(0)|\{C,Q\}|\Psi_0(0)\ra\nonumber\\
&=&\hbar^2\frac{E(0)}{\hbar\om_0}+\frac{\hbar^2}{\om_0^2}\dot{b}^2b^2\frac{E(0)}{\hbar\om_0}\left(1+\frac{E(0)}{\hbar\om_0}\right),
\eeqa
after using $\la\left\{Q,C\right\}\ra=0$.

\section{Driving protocols} 

In what follows, we analyze the exact nonadiabatic energy fluctuations for specific instances of driving protocols of the time-dependent harmonic frequency.

\subsection{Free expansion}
A standard protocol for time-of-flight imaging of ultracold atoms relies on suddenly releasing a cloud initially trapped in a harmonic potential with frequency $\om_0$. For $t>0$, $\om(t)=0$. For an initial equilibrium state at $t>0$, the initial conditions $b(0)=1$ and $\dot{b}(0)=0$ are required for the time-dependent state (\ref{psit}) to reduce to the initial state.
The solution of the Ermakov equation is then $b(t)=\sqrt{1+\om_0^2t^2}$ \cite{MinguzziGangardt05,delcampo08}. In this case there is no reference notion of adiabatic dynamics and $Q^*$ is not defined. The mean energy satisfies
\beqa
\frac{\la H(t)\ra_t}{\la H(0)\ra} =\left[\frac{1}{2b(t)^{2}}+\frac{\om(t)^2 b(t)^2}{2\om_0^2}+\frac{\dot{b}(t)^2}{2\om_0^2}\right]=1,
\eeqa
i.e, the mean total energy is preserved during time of flight.
The same holds true for  the expectation value of all the moments of the Hamiltonian. In turn,
for an energy eigenstate $\la H_0^2\ra=\la H_0\ra^2$ and thus $\la H_t^2\ra_t- \la H_t\ra_t^2=0$ for all $t\geq0$.

\subsection{Sudden quenches}
Consider the sudden modulation of the trap frequency so that it varies from an initial value $\om_0$ for $t<0$ to a final value $\om_F$.
Sudden quenches  of this type have been studied in a variety of applications such as probing quantum correlations in ultracold gases, the study of their relaxation and thermalization \cite{MinguzziGangardt05,delcampo08,Rajabpour14}, the analysis of Loschmidt echoes and work statistics \cite{DeffnerLutz08,Campbell14,GarciaMarch16,Vicari19}, 
and finite-time quantum thermodynamics of thermodynamic cycles \cite{Rezek06,Salamon09,Rezek09,Hoffmann11,Abah12,Jaramillo16,Kosloff17},  among other examples.

The solution of the Ermakov equation with $b(0)=1$ and $\dot{b}(0)=0$ leads to
the scaling factor that exhibits undamped oscillations  \cite{MinguzziGangardt05}
\beqa
b(t)=\sqrt{1+(\om_0^2/\om_F^2-1)\sin^2(\om_Ft)}.
\eeqa
The nonadiabatic factor $Q^*$ accounting for quantum friction reads \cite{Jaramillo16}
\beqa
Q^*(t)=\frac{\om_0^2+\om_F^2}{2\om_0\om_F}.
\eeqa
As a result, 
\beqa
\la H_t^2\ra_t- \la H_t\ra_t^2=\hbar^2\om_F^2\frac{E(0)}{\hbar\om_0}
 \frac{(\om_0^2-\om_F^2)^2}{2\om_0^2\om_F^2}.
\eeqa

\subsection{Linear frequency ramp}

\begin{figure}[t]
\begin{center}
\includegraphics[width=1\linewidth]{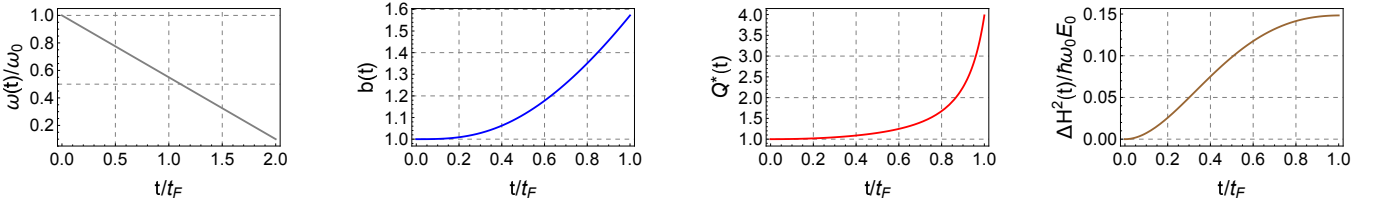}
\end{center}
\caption{\label{Fig3Lin} {\bf  Scale-invariant expansion induced by a linear frequency ramp.} A linear ramp of the trap frequency is considered with $\om_F=\om_0/10$ and $t_F=2/\om_0$. The scaling factor is derived by solving the Ermakov equation and is shown to exhibit a monotonic growth in this time scale. The nonadiabatic factor and the energy variance share the monotonic growth with the time of evolution.}
\end{figure}

As an example of a generic nonadiabatic driving we consider a linear ramp of the frequency such that $\om(t)=\om_0$ for $t<0$, $\om(t)=\om_0+(\om_f-\om_0)t/t_F$ for $0<t<t_F$ and $ \om(t)=\om_F$ for $t\geq t_F$.  Such time-dependence is nonanalytic at $t=\{0,t_F\}$ and is thus expected to give rise to  nonadiabatic dynamics \cite{Lidar09}. The limit of $t_F\rightarrow0$ naturally reduces to a sudden quench. The scaling factor can be obtained by solving numerically the Ermakov equation subject to $b(0)=1$ and $\dot{b}(0)=0$ and is shown  in Fig. \ref{Fig3Lin}. The nonadiabatic factor increases monotonically as a function of time with the final mean-energy being  four times higher than the corresponding adiabatic value.
The energy variance exhibits as well a monotonic behavior, with lower values  reflecting the expansion nature of the process, during which the instantaneous trapping frequency decreases. Such features are however dependent on the parameters chosen and a  non-monotonic behavior of $b(t)$, $Q^*$ and $\Delta H(t)^2$ is observed whenever $t_F$ is of the order of the inverse of the frequencies involved in the process, e.g., $\om_0^{-1}$ and  $\om_F^{-1}$.

We note that exact closed-form solutions of the Ermakov equation can be constructed as discussed in \cite{Kim16}.

%

\subsection{Shortcuts to adiabaticity by reverse engineering}
\begin{figure}[t]
\begin{center}
\includegraphics[width=1\linewidth]{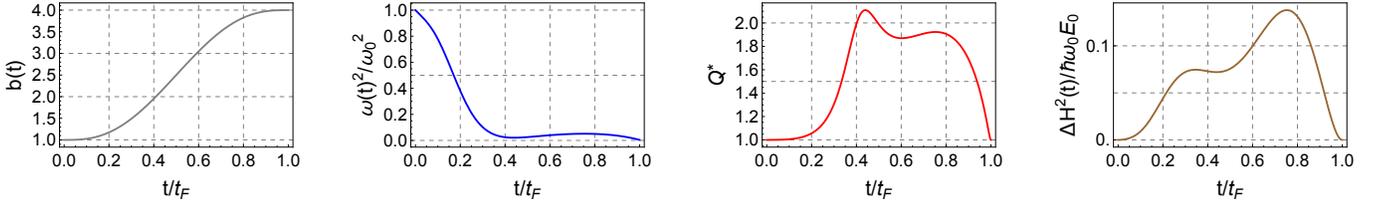}
\end{center}
\caption{\label{Fig1RevEng} {\bf Reverse engineering of the scale-invariant dynamics.} Given an interpolating scaling factor $b(t)$ as a function of time, the frequency modulation $\om(t)$  is obtained from the Ermakov equation.  The scaling factor, frequency square, nonadiabatic factor and energy fluctuations are shown from left to right  during a shortcut to adiabaticity for an expansion with final scaling factor $b(t_F)=4$ and duration $t_F=10/\om_0$. The latter is chosen long enough so that $\om^2(t)>0$ during the process. While values of the nonadiabatic factor $Q^*>1$ manifest the nonadiabatic nature of the process, the time-evolving state reduces to a stationary equilibrium state at the beginning and end of the protocol, when $Q^*=1$. The nonadiabatic energy variance, when normalized by $\hbar\om_0E(0)$, accounts for the behavior of the ground state of the harmonic oscillator, the Calogero-Sutherland gas, and the low-energy state of the unitary Fermi gas discussed. Its vanishing value at $t=\{0,t_F\}$  indicates that  the initial and final state are energy eigenstates of the corresponding Hamiltonians, $H(0)$ and $H(t)$, respectively.
}
\end{figure}

Reverse engineering provides a convenient approach for the fast driving of  a system from an initial equilibrium state towards a final nonequilibrium state, without relying on adiabatic strategies. 
Fixing the duration of the protocol to be $t_F$, with initial and final frequencies $\om_0$ and $\om_F$, we proceed to impose the following boundary conditions on the scaling factor: $b(0)=1$ and $b(t_F)=b_{\rm ad}(t_F)=\sqrt{\om_0/\om_F}$. Additional boundary conditions are required for the exact solution (\ref{psit}) to reduce to a stationary state. The vanishing of the derivates of the scaling factor $\dot{b}(0)=\dot{b}(t_F)=0$ make sure that $Q^*=1$ at $t=\{0,t_F\}$.  It proves convenient to further impose $\ddot{b}(0)=\ddot{b}(t_F)=0$. With these conditions one can then fix an interpolating polynomial ansatz for $b(t)$ and determine the driving protocol $\om(t)$, using the Ermakov equation as $\om(t)^2=\om_0/b^4-\ddot{b}/b$. This is the essence to design a shortcut to adiabaticity by reverse engineering the scale-invariant dynamics \cite{Chen10,delcampo11,Beau16}, as experimentally demonstrated in 
\cite{Schaff10,Schaff11,Deng18}. Taking the interpolating ansatz to be $b(t)=1+10(t/t_F)^3(b(t_F)-1)-15(t/t_F)^4(b(t_F)-1)+6(t/t_F)^5(b(t_F)-1)$, figure \ref{Fig1RevEng} shows the  modulation of the frequency-squared $\om(t)^2$, nonadiabatic factor $Q^*$ and nonadiabatic energy fluctuations for a protocol with $t_F=10/\om_0$ leading to an expansion with final $b(t_F)=4$.
The duration time is chosen such that $\om(t)^2\geq0$ for all $0\leq t\leq t_F$, preventing the inversion of the trap associated with values $\om(t)^2<0$ required for shorter processes in which $t_F\sim\om_0^{-1}$. The nonadiabatic nature of the protocol is manifested in the values of $Q^*$ exceeding unity for $0< t< t_F$. Nonetheless, $Q^*=1$ and $\Delta H(t)^2=0$ for $t=\{0,t_F\}$ showing that at the beginning and end of the protocol excitations are cancelled out and the nonadiabatic evolution reduces to a stationary energy eigenstate.
%

\subsection{Local counterdiabatic driving}

\begin{figure}[t]
\begin{center}
\includegraphics[width=1\linewidth]{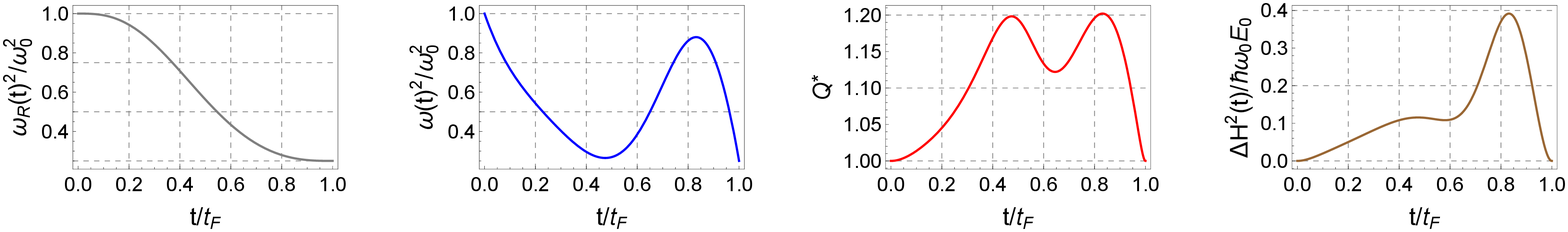}
\end{center}
\caption{\label{Fig2LCD} {\bf Local counterdiabatic driving of the scale-invariant dynamics.} Given a reference modulation of the  trapping frequency $\om_R(t)$  between the  initial and final trap configurations, a shortcut to adiabaticity by local counterdiabatic driving implements a modulation of the trapping frequency $\om(t)$  that differs from the reference one $\om_R(t)$.  The reference  modulation of the  frequency square,  the  frequency square under the local counterdiabatic driving, the corresponding nonadiabatic factor and energy fluctuations are shown from left to right  for an expansion with final frequency $\om_F=\om_0/2$ and duration $t_F=2/\om_0$. The latter is chosen long enough so that $\om^2(t)>0$ during the process. 
}
\end{figure}

Consider an arbitrary  modulation of the trapping frequency $\om_R(t)$ as a reference, interpolating between  a prescheduled initial value $\om_R(0)=\om_0$ and  a final one $\om_R(t_F)=\om_F$. 
Such modulation is generally nonadiabatic. Shortcuts to adiabaticity engineered by local counterdiabatic driving \cite{delcampo13,Beau16,Deng18,Deng18Sci,Diao18} suppress nonadiabatic excitations by implementing instead the driving protocol with frequency $\om(t)$, where 
\beqa
\label{om2lcd}
\omega(t)^2=\om_R(t)^2-\frac{3}{4}\left(\frac{\dot{\om}_R(t)}{\om_R(t)}\right)^2+\frac{1}{2}\frac{\ddot{\om}_R(t)}{\om_R(t)}.
\eeqa
Driving the system with $\om(t)$, it follows from the Ermakov equation (\ref{EPE}) that the scaling factor is precisely 
\beqa
\label{blcd}
b(t)=\sqrt{\om_0/\om_R(t)}.
\eeqa 
With the boundary conditions, $\dot{\om}_R(0)=\ddot{\om}_R(0)=0$ and $\dot{\om}_R(t_F)=\ddot{\om}_R(t_F)=0$, 
we choose the interpolating ansatz  $\om_R(t)=\om_0+10(t/t_F)^3(\om_F-\om_0)-15(t/t_F)^4(\om_F-\om_0)+6(t/t_F)^5(\om_F-\om_0)$.
It is then guaranteed that $\omega(0)=\omega_R(0)=\om_0$ and $\omega(t_F)=\omega_R(t_F)=\om_F$, so the  local counterdiabatic protocol $\om(t)$ still interpolates between the same initial and final frequencies considered in the reference protocol $\om_R(t)$. Using Eqs. (\ref{om2lcd}) and (\ref{blcd}), the nonadiabatic factor (\ref{Qstar})  reads in this case
\beqa
Q^*=\frac{\om_R(t)}{\om(t)}\left[1+\frac{1}{4}\left(\frac{\ddot{\om}_R(t)}{\om_R(t)^3}-\frac{\dot{\om}_R(t)^2}{\om_R(t)^4}\right)\right].
\eeqa
This result corrects the expression found in \cite{Beau16,AbahLutz17}. 
The nonadiabatic energy variance takes the form
\beqa
\la H_t^2\ra_t- \la H_t\ra_t^2=\hbar^2\omega(t)^2\frac{E(0)}{\hbar\om_0}\left[\left(Q^*\right)^2-1\right],
\eeqa
and vanishes identically at $t=\{0,t_F\}$. Figure \ref{Fig2LCD} shows a polynomial modulation of the reference frequency-square $\om_R^2(t)$ that decreases monotonically during the process from $\om_0$ to $\om_F$. By contrast,  the modulation of $\omega(t)^2$ associated with the local counterdiabatic driving protocol is highly non-monotonic, involving a fast expansion, followed by a compression and an expansion again. This specific modulation as a function of time  strongly depends on the duration of the process and is shown here for $t_F=2/\om_0$ and an expansion with $\om_F=\om_0/2$.  The corresponding nonadiabatic factor $Q^*$ is shown to be greater than unity but at the beginning and end of the protocol when excitations are cancelled. The nonadiabatic energy variance grows from zero value  and fluctuates during the process until its completion, when it vanishes again, as the state at $t=t_F$ is an eigenstate of the final Hamiltonian.

%

\section{Conclusions and Discussion}

Scale invariance is an important symmetry that governs the dynamics of a variety of systems, including many harmonically-trapped ultracold atomic gases. It is not only of relevance to time-of-flight imaging techniques, but it also plays an important role in the implementation of quantum control techniques and the characterization of  finite-time thermodynamics. We have shown that nonadiabatic energy fluctuations in many-body systems under scale-invariant driving can be simply expressed in terms of the equilibrium properties of the initial state. In particular, knowledge of expectation value of the second moments of the initial Hamiltonian, square position and squeezing operators suffices to characterize the exact nonadiabatic dynamics in terms of the evolution of the scaling factor, which governs the size of the atomic cloud. As the later is an experimentally measurable quantity, our results provide a way to characterize nonadiabatic energy fluctuations in the laboratory using time-of-flight imaging techniques. To illustrate our results, we have  considered the single-particle quantum oscillator of relevance to experiments with trapped ions and thermal atomic clouds, as well as two instances of many-body systems: the Calogero-Sutherland gas and the unitary Fermi gas. The corresponding nonadiabatic evolution has been characterized in expansion protocols involving sudden quenches and finite-time protocols, including linear frequency ramps, as well as shortcuts to adiabaticity leading to stationary states upon completion of the process.

The exact results for the nonadiabatic energy fluctuations we have provided are expected to have widespread applications in both theoretical and experimental studies of these systems.  For example, they can be used to characterize far-from-equilibrium dynamics in ultracold gases, quantify the cost of quantum control  such as  the use of optimal protocols \cite{Salamon09,Stefanatos10,Stefanatos17,Larocca20} and shortcuts to adiabaticity \cite{Demirplak2008,Campbell17}, characterize the performance of devices and processes in quantum thermodynamics \cite{delCampo2018}, and study the ultimate speed limits \cite{MT45,Shanahan18}, quantum decay \cite{delcampo16} and orthogonality catastrophe \cite{Fogarty20} under scale-invariant quantum dynamics.

\section*{Acknowledgements}
It is a pleasure to acknowledge  discussions  with  Juan Jaramillo and Doha Mesnaoui, comments on the manuscript by Fernando J. G\'omez-Ruiz,  and insightful clarifications by I\~nigo L. Egusquiza. This research was partially funded by the John Templeton Foundation.

\bibliography{EnergyFluct_lib}

\end{document}